# Signature of charge-3*e* tunneling in twisted cuprate Josephson junctions

Yuying Zhu[1,6#*], Heng Wang[2,3,4#], Zechao Wang[3], Zhiyuan Sun[2,5],

Qi-Kun Xue[1,2,3,4,5*], and Ding Zhang[1,2,5,6*]

[1]Beijing Academy of Quantum Information Sciences, Beijing 100193, China

[2]State Key Laboratory of Low Dimensional Quantum Physics and Department of Physics, Tsinghua University, Beijing 100084, China

[3]State Key Laboratory of Quantum Functional Materials, Department of Physics, and Guangdong Basic Research Center of Excellence for Quantum Science, Southern University of Science and Technology, Shenzhen 518055, China

[4]Quantum Science Center of Guangdong-Hong Kong-Macao Greater Bay Area, Shenzhen 518045, China

[5]Frontier Science Center for Quantum Information, Beijing 100193, China

[6]Hefei National Laboratory, Hefei 230088, China

# These authors contributed equally.

*Email: zhuyy@baqis.ac.cn

qkxue@mail.tsinghua.edu.cn

dingzhang@mail.tsinghua.edu.cn

# Abstract

Cooper pairing is a fundamental aspect of superconductivity and results in charge transport with a basic unit of $2e$ ($e$ is the electron charge). This basic unit can manifest itself as the denominator in the formula for Shapiro steps of ac Josephson effect—$nhf/(2e)$, where $h$ is the Planck constant, $f$ is the ac frequency and $n$ is an integer. Violation of this standard formula may suggest unconventional physics, such as $nhf/4e$ for charge-$4e$ superconductivity. Here, we report discovery of a new sequence of $nhf/3e$ in twisted cuprate Josephson junctions. The microwave frequency/power, temperature, and current annealing dependences of this exotic sequence consistently point to $hf/3e$ as the unit for Shapiro steps, indicating a Josephson tunneling process involving a trio of electrons ($3e$). This bunching of an odd number of charges distinguishes itself from well-established superconducting processes, which are carried by composite bosons made of an even number of charges. The unusual observations highlight the important interplay between competing charge orders in twisted bilayer systems.

## Introduction

The Josephson effect is a zero-energy quantum tunneling process that has profound impact on both fundamental research[1-4] and frontier applications[5-8]. The tunneling current $I$ between the two superconductors is governed by the first Josephson equation[9] of $I = I_c \, sin \, \varphi$, where $I_c$ is the critical current and $\varphi$ is the phase difference between the two superconductors across the junction. The phase difference can be driven into rapid variation by a bias voltage $V$, following the second Josephson equation: $\frac{d\varphi}{dt} = \frac{2eV}{\hbar}$. Here, $2e$ reflects the fact that each tunneling entity consists of a pair of electrons—as was proposed by Leon Cooper in 1956[10]. In practice, the applied voltage $V$ is a few tens of microvolts such that the Josephson current oscillates at gigahertz frequencies (1 GHz~2 μV). Microwave irradiation can demodulate this ac Josephson effect, giving rise to current-voltage characteristics with current plateaus—Shapiro steps—at quantized voltage values of $nhf/2e$[11]. Notably, charge-$2e$ in the denominator reflects the tunneling of Cooper pairs[10].

It was soon realized that the simple harmonic of the first Josephson equation or the current-phase relation (CPR) only works for an ideal superconductor-insulator-superconductor (SIS) junction. A superconductor-normal metal-superconductor (SNS) junction, for instance, may possess different transparency of the normal metal section, such that the CPR deviates from the perfect sinusoidal form[9]. In the limit of a highly transparent junction, the CPR approaches $I = I_c \, sin \frac{\varphi}{2}$ such that microwave irradiations give rise to Shapiro steps solely at even integer numbers of $V_0$: $hf/e$, $2hf/e$, ... The missing of odd integral Shapiro steps was predicted to be a feature of topological superconductivity too[12-14]. By taking into account of the transparency of the tunneling barrier, a general form of the CPR reads: $I = \sum I_{cm} \, sin(\mathrm{m}\varphi)$, where $m$ is an integer and $m = 2, 3, \ldots$ represent higher order terms ($I_{cm}$ is the corresponding coefficient). Of late, it was demonstrated that a SIS junction can also host higher harmonics due to the existence of multiple tunneling channels[15] or co-tunneling of Cooper pairs[16,17]. In both SNS and SIS junctions, the higher harmonics ($m \geq 2$) manifest themselves as Shapiro steps at $nhf/(2me)$ in the ac Josephson effect. For example, if $I = I_{c1} \sin\varphi + I_{c2} \sin(2\varphi)$, Shapiro steps would occur at half integers ($\pm 1/2$, $\pm 3/2$, …), in addition to the main sequence at $n = \pm 1$, $\pm 2$,…[18-20], with the voltage unit of $V_0 = hf/2e$.

Here, we report Shapiro steps with a bizarre sequence of $\pm hf/3e$, $\pm 2hf/3e$, $\pm hf/e$, … across a wide temperature range (up to 70 K) in two Josephson junctions made of $Bi_2Sr_2CaCu_2O_{8+x}$ (BSCCO) flakes stacked along the *c*-axis with a twist angle of 45 degrees (Fig. 1a). The unconventional sequence draws a distinct contrast to the standard sequence reported before. Both the microwave power and measurement temperature dependences of the unconventional sequence suggest that the basic voltage unit for the Shapiro steps now becomes $hf/3e$ instead of $hf/2e$. Furthermore, asymmetrically distributed Shapiro steps and half-integer Shapiro steps emerge with the revised voltage unit ($hf/3e$) after current annealing the same sample. These surprising outcomes indicate a non-trivial Josephson tunneling process with an effective charge of $3e$. It stands in sharp contrast to the conventional picture of superconductivity, which relies on an even number of charge carriers. Our results underscore the fertile interplay between twistronics and high-temperature superconductivity.

## Results

### Unconventional sequence of Shapiro steps

Figure 1a is a schematic drawing of our twisted cuprate junction, consisting of two mechanically exfoliated flakes of BSCCO. It is fabricated by taking advantage of the cryogenic van der Waals stacking method, as described elsewhere[21-23]. Notably, the top cuprate block is twisted around the *c*-axis by 45 degrees relative to the bottom one, as confirmed by high-resolution transmission electron microscopy study in Extended Data Fig. 1. We are able to measure exclusively the Josephson tunneling at the twisted interface, as illustrated by the downward arrow, owning to the twist and the relatively small overlapping area between the top and bottom blocks[21-23].

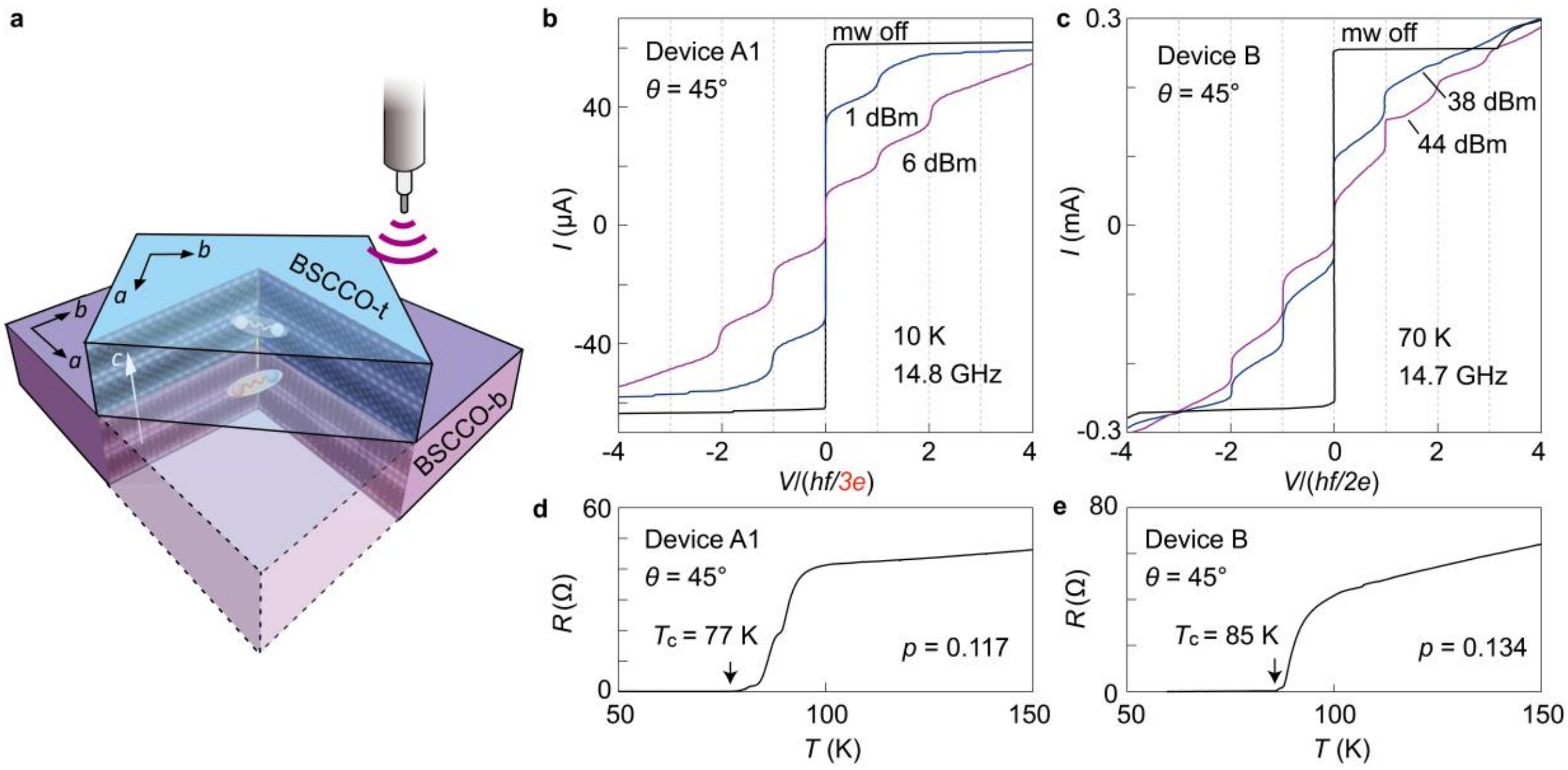

***Figure 1: Charge-3e and charge-2e Josephson tunneling. a,*** *Schematic drawing of a 45°-twisted cuprate Josephson junction cut open with two orthogonal cross-sectional views (based on transmission electron microscopy data). Microwave irradiation is realized via sending the high frequency signal to an open-ended coaxial cable positioned close to the sample.* ***b, c,*** $I$-$V$ *characteristics of device A1 and device B with or without microwave irradiation. Numbers in the panels mark the twist angles, measurement temperatures, microwave frequencies, and microwave powers. Abscissa of panel* ***b*** *or* ***c*** *is plotted with the quantized unit of* $hf/3e$ *or* $hf/2e$*.* ***d, e,*** *Temperature dependence of the junction resistances for the two samples. Arrows mark the superconducting transition temperature* $T_c$*. Doping levels p are obtained from* $T_c$ *by using the empirical relation:* $T_c = 91 \cdot [1 - 82.6(p - 0.16)^2]$.

Figure 1b and c draw the sharp contrast between the ac Josephson effects of two samples: device A1—focus of this study—and device B—the control sample. We could reproduce the unconventional ac Josephson effect in another sample—device A2 (Extended Data Fig. 2). Under microwave irradiations, both device A1 and device B show Shapiro steps at discrete voltage values with equal separations. While the Shapiro steps in Fig. 1c take the standard unit of $hf/2e$, those steps in Fig. 1b possess an unusual unit of $hf/3e$, which has never been observed before. The contrast in the quantized units epitomizes our key finding: device A1 hosts an anomalous sequence of Shapiro steps, other than the standard sequence of $nhf/2e$. Notably, we employ the same measurement setup for device A1 and device B. Instrumental uncertainty, such as inaccurate microwave frequency, can be ruled out as the origin for the unusual results.

Apart from the different quantized units for ac Josephson effect, both devices A1 and A2 (Fig. 1d, Extended Data Fig. 2b) exhibit superconducting transition temperatures $T_c$ of about 10 K lower than the optimal $T_c$ of 88 K in device B (Inset of Fig. 1e). It indicates that the unconventional Shapiro steps appear in the underdoped regime of superconductivity in the twisted junction. Oxygen out-diffusion[24] during the cleave-and-stack process seems to play an essential role in forming such unique junctions.

**Microwave frequency and power dependence**

In Fig. 2, we corroborate the robustness of the unusual ac Josephson effect by plotting $dV/dI$ (Fig. 2a, b, c) or $dI/dV$ (Fig.2d, e, f) as a function of the bias current or voltage and the microwave power $P$, at selected microwave frequencies. The pajama-like pattern in Fig. 2a, b, c reflects a series of Shapiro steps (dark-blue regions with non-zero numbers). The vertical expansion of each dark-blue region corresponds to the width of the Shapiro plateau at a fixed $P$. Increasing the microwave power can periodically modulate the width of the Shapiro steps. The steps first become more prominent and then get suppressed with increasing $P$. They re-emerge after further increasing $P$ but the extension in $I$ gets reduced. This process then repeats itself at higher values of $P$, similar to the Fraunhofer pattern. Plotting the same data set but as a function of voltage and microwave power (Fig. 2d, e, f) better showcases the quantized voltage positions of Shapiro steps at $nhf/3e$ with $n$ extending from $\pm 1$ to $\pm 7$.

To quantitatively extract the voltage positions, we sum up $dI/dV$ at each $V$ over the microwave power from 0 to 15 dBm. Figure 2g shows a section of the summed-up data at 10 K, 14.8 GHz. It is noteworthy that the sharp peaks of $\Sigma dI/dV$, especially at $hf/3e$, $2hf/3e$, possess a prominent voltage separation. If there would exist an additional Shapiro step, it would correspond to a peak at $hf/2e$ (dotted curve), which should be clearly observable by experiment and would not be masked by the peaks at $hf/3e$ and $2hf/3e$. Therefore, the absence of a standard sequence of Shapiro steps is not due to limited resolution of our experiment. Figure 2h summarizes the frequency dependence of Shapiro steps (voltage positions of the peaks in the summed-up data, such as those in Fig. 2g) on both the positive and negative sides at 10 K. All the voltage values follow the expected linear behaviors with frequency (dashed lines), although the prefactor now becomes $nh/3e$.

In Fig. 2i, we extract the width of the current plateaus $\Delta I_n$ at $nhf/3e$, with $n = 0, 1, 2, 3, 4$, and plots them as a function of $P$. The power modulation follows the expected behavior with a monotonically increasing $n$. Namely, the local maxima(minima) of $\Delta I_{n+1}$ occur at around the values of $P$ for the local minima(maxima) of $\Delta I_n$. And the plateau width decreases with increasing $n$: $\Delta I_{n,max} > \Delta I_{n+1,max}$. For example, the plateau width for $n = 1$ can reach 10 μA but the maximum plateau widths for $n = 2$ and 3 become about 7 μA and 2 μA, respectively. This monotonic decreasing behavior speaks against a competing scenario to explain the unusual sequence. Namely, the steps at $hf/e$ and $2hf/e$ could be assigned to the main sequence of a fully transparent standard Josephson junction whereas steps at $hf/3e$, $2hf/3e$, etc. would correspond to fractional Shapiro steps. In this competing scenario, the plateau widths of $hf/e$ and $2hf/e$, the integral Shapiro steps, would be much more pronounced than the fractional steps at $hf/3e$ and $2hf/3e$. This would give rise to oscillatory strength of $\Delta I_{n,max}$ with $n$, instead of the experimentally observed monotonic decrease with $n$.

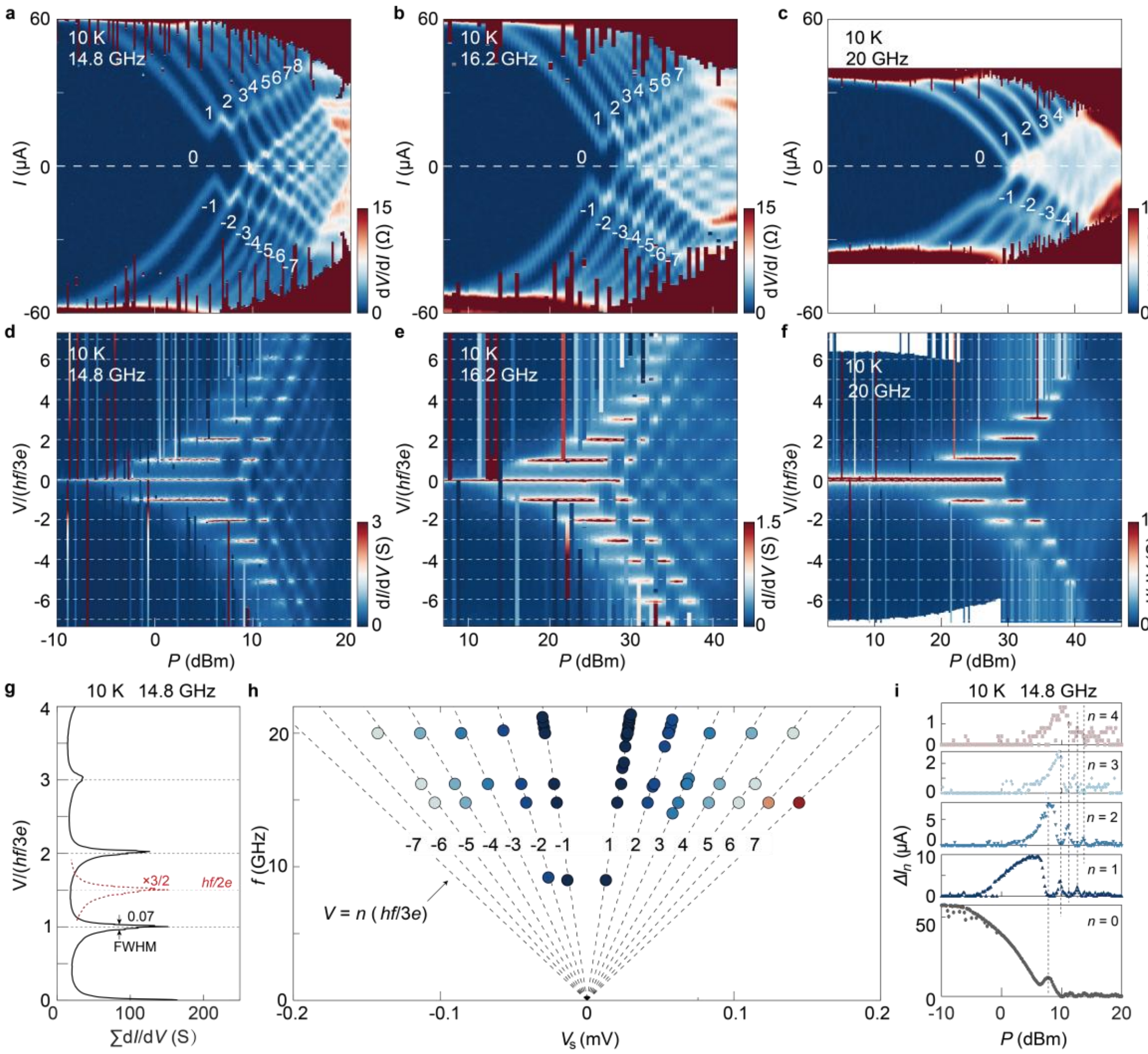

***Figure 2: Frequency dependent studies of charge-3e Josephson tunneling. a, b, c,*** *$dV/dI$ across the twisted junction of device A1 as a function of bias current $I$ and microwave power $P$ at three different frequencies. Non-zero integral numbers indicate the Shapiro steps.* ***d, e, f,*** *Same data set of* ***a, b, c*** *but plotted as $dI/dV$ versus $V$ and $P$.* ***g,*** *Summation of $dI/dV$ over the microwave power (0 to 15 dBm, except for data with erratic jumps) plotted as a function of $V$. Dotted curve indicates a peak at $hf/2e$ (by scaling the original data around $hf/3e$ with a factor of 3/2).* ***h****, Voltage positions of the Shapiro steps as a function of frequency $f$ for $n = \pm 1, \ldots, \pm 7$. At each frequency, we sum up the data of $dI/dV(V)$ at different $P$ (panel* ***g****), at which the data hosts no erratic jumps, and then take the peak positions as $V_s$. Dashed lines indicate the theoretical relation: $V_s = nhf/3e$. Error bars, defined as the full-width-at-half-maximum (FWHM) of the peaks, are smaller than the symbol size here.* ***i,*** *Microwave power dependence of the plateau width ($\Delta I_n$) for the zero-bias critical current ($n = 0$) and first four Shapiro steps (n= 1-4) based on the data in panel* ***a****. $\Delta I_n$ is defined as the current span where $|V - V_S(n)| < 10\%$ of $V_S(n)$.*

**Temperature dependence**

In Fig. 3a, 3b and 3c, we show the typical data obtained at elevated temperatures of 30 K, 50 K, and 70 K. It indicates that the unconventional sequence of Shapiro steps is robust across a wide temperature range. In Fig. 3d, we show both the critical current (without microwave irradiation) and the extracted voltage positions of Shapiro steps as a function of temperature. The critical current (top panel) follows the standard temperature dependence and diminishes at around 80 K. Meanwhile, the unusual sequence of Shapiro steps persists from 1.6 K up to 70 K, covering a wide range: [$2\%T_c$, $93\%T_c$ ]. This persistence further rules out the possibility that the observed unconventional ac Josephson effect consists of a mixture of integral and fractional Shapiro steps, because the fractional Shapiro steps would vanish more quickly with increasing temperature than the integral steps do. Figure 3e and Extended Data Figure 3 confirm that the exotic results of device A1 are reproducible in multiple thermal cycles, distinguishing themselves from the metastable fractional steps due to trapped fluxes[25,26].

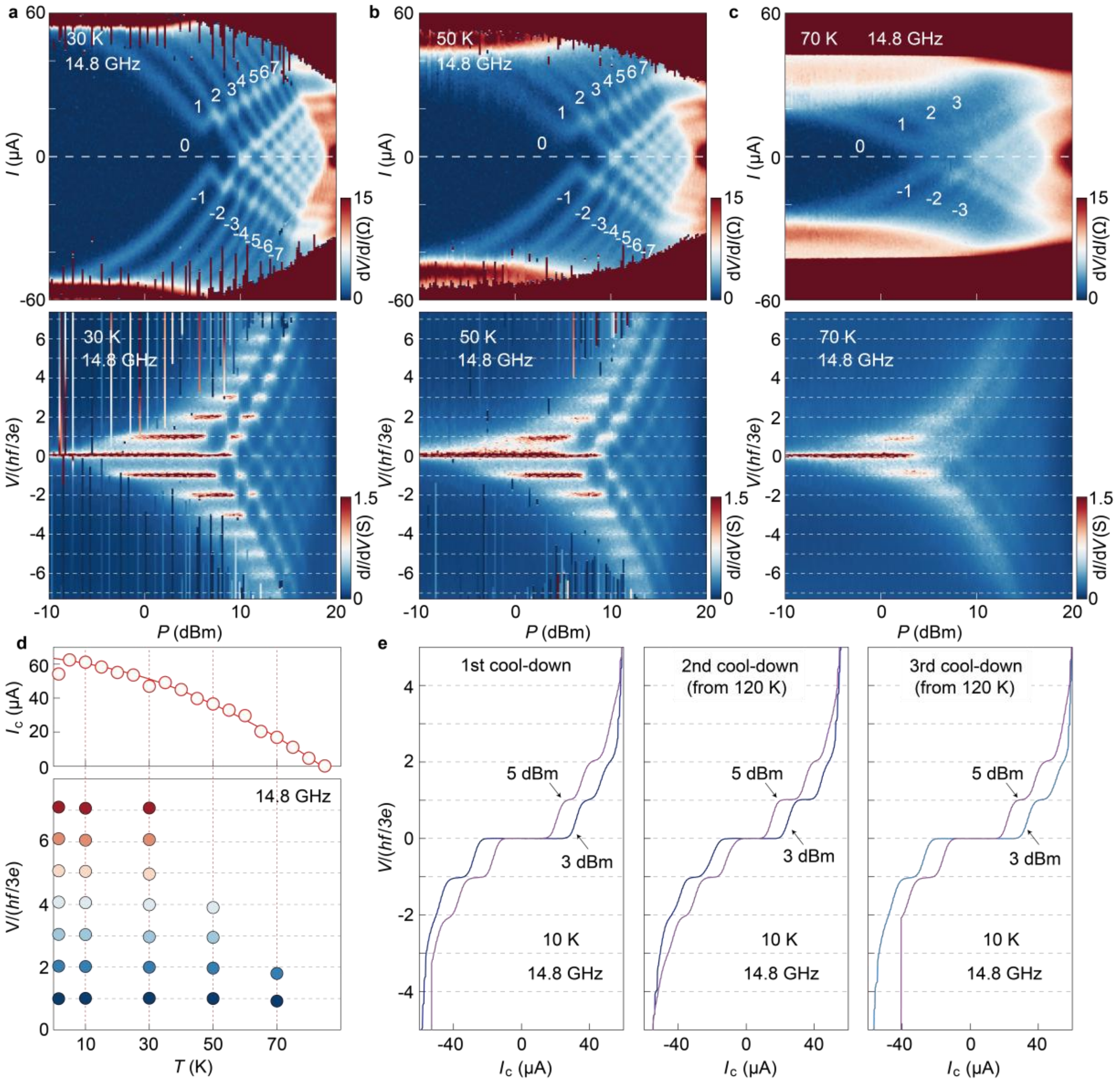

***Figure 3: Temperature dependent studies of charge-3e Josephson tunneling. a, b, c,*** *Top: $dV/dI$ across the junction as a function of bias current $I$ and power P at a fixed microwave frequency and temperature (30 K, 50 K, 70 K). Bottom: corresponding $dI/dV$ versus $V$ and $P$.* ***d,*** *Top: Temperature dependence of the critical current $I_c$ of device A1 without microwave irradiations. Bottom: Voltage positions of Shapiro steps measured at a set of temperature points. Error bars for $V_S$ are smaller than the symbol size.* ***e,*** *Typical $I$ - $V$ characteristics under microwave irradiation (14.8 GHz, 3 dBm and 5 dBm) in three separate cool-downs.*

**Superconducting diode effect and fractional Shapiro steps**

Despite the unusual voltage quantization unit of $hf/3e$, device A1 exhibits a similar response to current annealing as the standard twisted cuprate junction does[27]. Namely, device A1 can show asymmetric transport—superconducting diode effect (SDE)—after current annealing. Figure 4a illustrates the annealing protocol. We cool the sample from 150 K, well above $T_c$, to 10 K while applying a relatively large dc current (1 to 2 mA) through the junction. After withdrawing the large current at 10 K, we carry out the measurement of the ac Josephson effect. Figure 4b shows the exemplary $I$-$V$ characteristics with pronounced and stable asymmetry between the branches in the positive and negative current directions. Microwave irradiation dramatically enhances the asymmetry and gives rise to quantum superconducting diode effect (QSDE) (purple and pink curves of Fig. 4b). The asymmetric Shapiro steps are still quantized with the revised unit of $hf/3e$. Figure 4c and Extended Data Figure 4 demonstrate the stable operation of QSDE. Under a square wave of input current, the output voltage alternates between zero and a quantized value—$hf/3e$ or $2hf/3e$ and can be switched off by turning off the microwave. Similar to our previous works[25,27], we can recover symmetric $I$-$V$ characteristics in device A1 by thermal cycling to a temperature above $T_c$.

Apart from QSDE, the current annealing can also induce fractional Shapiro steps (Fig. 4d,e,f and Extended Data Fig. 5). Figure 4d shows that a clear fractional step at $hf/6e$ emerges in the $I$-$V$ characteristic of device A1 after current annealing at 1.5 mA. This emergent plateau corresponds to a half integer Shapiro step in the new unit of $hf/3e$. Signatures of fractions at $3/2$ and $5/2$ of $hf/3e$ can also be seen, as indicated by the colored plots of Fig. 4e and 4f. The observation of annealing induced half-integer steps reaffirms that the Shapiro steps before annealing (data in Fig. 2, for example) belong to the integral category.

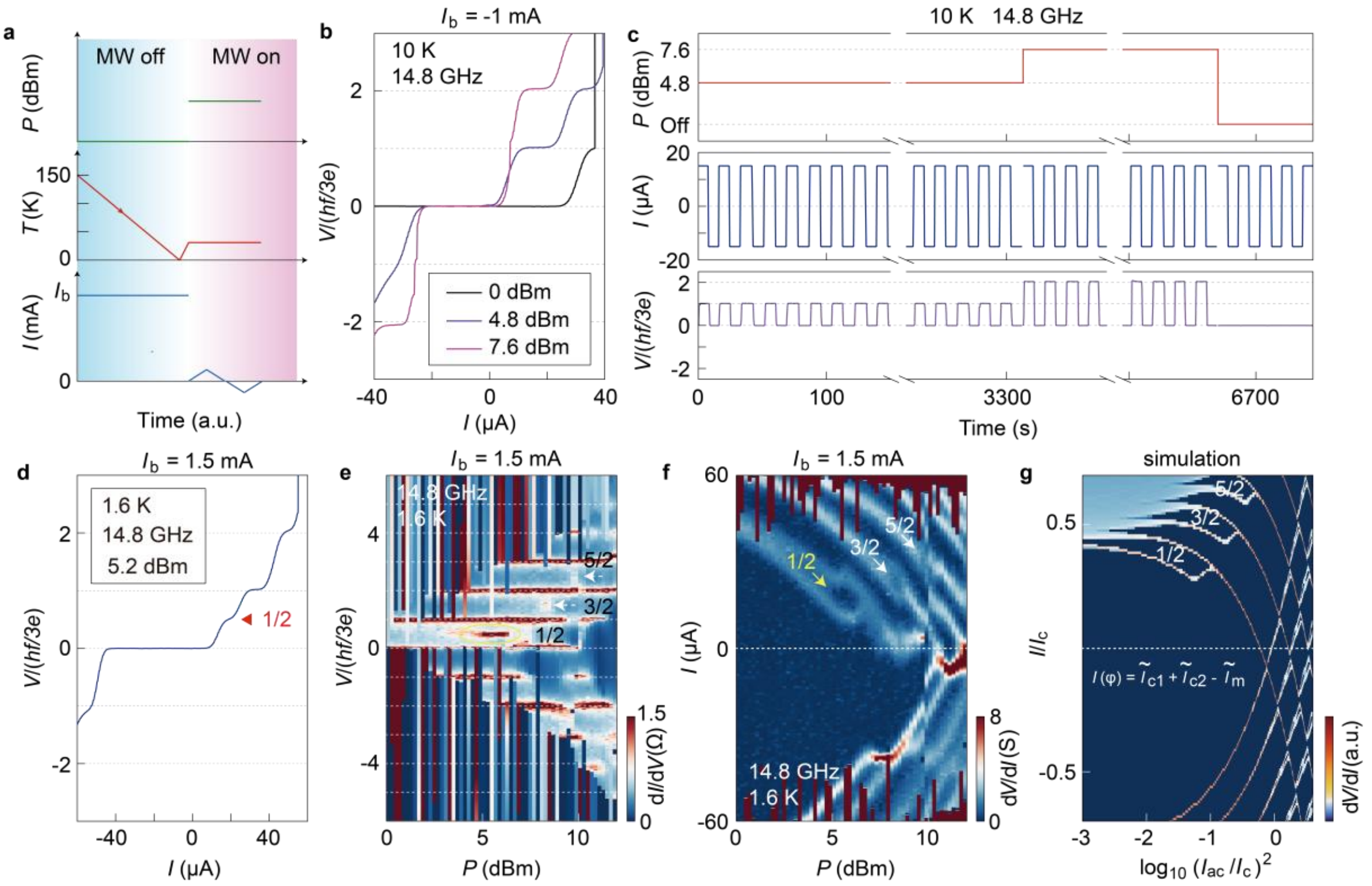

***Figure 4: Quantum superconducting diode effect and fractional Shapiro steps. a,*** *Illustration of the current annealing protocol. Here,* $I_b$ *(about 1-2 mA) denotes the bias current applied across the junction.* ***b,*** $I$ - $V$ *characteristics indicating superconducting diode effect and quantum superconducting diode effect of device A1. Data are obtained after current annealing with* $I_b$ *= -1 mA.* ***c,*** *Half-wave rectification with quantized output. Top to bottom: input microwave power, input current, and output voltage as a function of time.* ***d,*** $I$-$V$ *characteristic of device A1 after current annealing with* $I_b$*= 1.5 mA and under microwave irradiation (f = 14.8 GHz, P = 5.2 dBm). Red arrow indicates the half-integer Shapiro step at 1/2 (in unit of* $hf/3e$*).* ***e,*** $dI/dV$ *versus* $V$ *and* $P$ *after current annealing with* $I_b$*= 1.5 mA. White arrows indicate features at 3/2 and 5/2.* ***f,*** $dV/dI$ *across the junction as a function of* $I$ *and P. Yellow arrow highlights the fractional Shapiro step at 1/2. White arrows mark the regions that indicate fractional Shapiro steps at 3/2 and 5/2.* ***g,*** *Theoretical modelling of asymmetrically developed Shapiro steps. We include* $\tilde{I}_{c1} = I_{c1}\,sin(3eVt/\hbar)$, $\tilde{I}_{c2} = I_{c2}\,sin(6eVt/\hbar)$, *and* $\tilde{I}_m = I_m\,cos(3eVt/\hbar)$*. Parameters* $I_{c1}$, $I_{c2}$, *and* $I_m$ *are adjusted to simulate the experiment pattern in* ***f****. Numbers indicate the half integer steps (in unit of* $hf/3e$*).*

## Discussions

Our twisted junction is electrically connected to intrinsic Josephson junctions of the constituent BSCCO flakes. However, a network of standard Josephson junctions is unlikely to generate the unusual sequence of Shapiro steps, because the lowest possible step location for the total voltage cannot be smaller than $hf/2e$—the smallest step size for the standard sequence of Shapiro steps. This smallest step size already exceeds the first experimentally observed step at $hf/3e$. In a more exotic scenario, Shapiro steps of $nhf/3e$ may appear from two charge-6$e$ junctions connected in series—each junction hosts an effective tunneling of charge-6$e$. However, this situation would require the two junctions to have exactly the same characteristics, which is unlikely because experimentally there exists only one junction with two twisted copper-oxide planes. Furthermore, charge-6$e$ superconductivity usually occurs in a narrow temperature window where superconducting fluctuations in multiband systems play an important role[28]. By contrast, we observe $nhf/3e$ across a wide temperature range.

As alluded to, standard Shapiro steps at $nhf/2e$ reflect the ac Josephson effect following the CPR $I = I_c \, sin\, \varphi$ and the second Josephson equation of $d\varphi/dt = 2eV/\hbar$. Combining the two standard equations yields $I = I_c \, sin(2eVt/\hbar)$ (we assume $\varphi = 0$ at $t = 0$) with $2eV/h$ being the fundamental frequency. The experimental observation of unconventional Shapiro steps solely at $nhf/3e$ thus corresponds to a drastically modified time dependent equation: $I = I_c \, sin(3eVt/\hbar)$. In fact, Figure 4g shows that the emergence of SDE together with half-integer Shapiro steps (Fig. 4f) can be captured by further including the following terms: $I_{c2} \, sin(6eVt/\hbar)$ and $I_m \, cos(3eVt/\hbar)$. Here, the first term is the second harmonic and the second one represents time reversal symmetry breaking (TRSB). These two additional terms likely arise from trapped fluxes induced by current annealing, similar to our observation in normal Josephson junctions[25, 27, 29]. In addition, we note that introducing the conventional second harmonic and TRSB terms [$I_{c2} \, sin(4eVt/\hbar)$ and $I_m \, cos(2eVt/\hbar)$] would yield simulation that is incompatible with the experimental observation (Extended Data Fig. 6). It reaffirms the absence of the conventional term of $sin(2eVt/\hbar)$ as the fundamental harmonic here.

In general, the robust occurrence of $I = I_c \, sin(3eVt/\hbar)$ in multiple devices (A1 and A2) suggests fundamentally different physics than that in standard Josephson junctions hosting $I = I_c \, sin(2eVt/\hbar)$. This exotic behavior seems to require coherent tunneling of a trio of electrons (3*e*) across the weak link. Such an unprecedented phenomenon challenges the present understanding of twisted cuprate systems. In particular, the theoretically proposed topological superconductivity in twisted cuprates[17] has co-tunneling of Cooper pairs (4*e*) as the lowest order process, which cannot explain our observation. Here, we speculate on the microscopic mechanism by considering interlayer excitonic order[30]. Our device is in essence a bilayer system with relatively narrow bands and strong electron-electron interactions both within the copper-oxide planes and across the SrO/BiO/BiO/SrO barrier. Similar to a semiconducting bilayer[30-32] and graphene bilayers[33], the interlayer interaction may favor the binding of electrons in one layer with holes in the opposite layer, forming a state with interlayer excitonic order. The coexistence of high-temperature superconductivity with such an excitonic order might give rise to an effective tunneling process involving a Cooper pair and a single electron, giving rise to the fractional CPR: $sin(3\varphi/2)$. Together with the second Josephson equation, it would give rise to a time dependent term proportional to $sin(3eVt/\hbar)$. We point out that the fractional CPR ($sin(3\varphi/2)$) suggests exotic physics, because the energy of the junction does not go to itself when the superconducting phase difference increases by $2\pi$. However, this term would in general coexist with more fundamental terms such as $sin(\varphi/2)$ and $sin\,\varphi$. The reason for the sole appearance of $sin(3\varphi/2)$ calls for further investigations.

Experimentally, we point out that our twisted cuprate junctions (A1 and A2) possess a slight reduction in doping from the optimal regime[22], as shown in Extended Data Tab.

1. Josephson tunneling between twisted cuprate flakes in this slightly underdoped regime seems unique. By contrast, we have investigated another 12 samples (apart from the eight previously studied samples[25,27]) close to optimal doping (Extended Data Tab. 2) and found only the standard sequence of Shapiro steps there. The occurrence of unconventional Shapiro steps only in the underdoped regime therefore indicates the importance of fine-tuning the interplay between superconductivity and other competing orders such as charge ordering, which strengthens as the doping level $p$ approaches 0.12[34]. The sensitivity on doping can inspire similar studies on other twisted bilayer systems[35,36], where the band filling and the strength of electron-electron interactions are individually adjustable. The possible involvement of excitons as the underlying mechanism also adds a fresh perspective to the Coulomb interactions in bilayer systems[31-33]. Finally, our experiments show that extending to the microwave regime is beneficial for identifying unusual phenomena that are not attainable in standard quasi-dc transport.

## Methods

### Sample fabrication and instrument information

Sample fabrications follow the procedure described in our previous works[26]. Transport measurements were performed in a vibration-free closed-cycle system equipped with a superconducting magnet. To avoid remanent fields due to trapped vortices in the superconducting coil, we intentionally warmed up the system to a temperature of 300 K and cooled it down. In the following measurements, no current was sourced to the superconducting magnet. The microwave experiment was carried out by introducing the high frequency signal into the cryogenic system via a semi-rigid coaxial cable, which is open ended at the low temperature stage with the extended inner conductor serving as the antenna. The microwave signal is provided by the signal generator with a maximum frequency of 40 GHz or 20 GHz and a maximum output power of 30 dBm. For measurements with the microwave power up to 40 dBm, we employed the power amplifier. All microwave power values indicated in the figures represent the readings at the microwave source.

## Data availability

The source data for generating figures in the main text of this study have been deposited in: https://doi.org/ (will be inserted after the acceptance of the paper).

## Acknowledgements

We thank Genda Gu for providing the crystals. This work is financially supported by the Ministry of Science and Technology of China [Grant No. 2024YFA1408101 (YZ, HW), 2022YFA1403100 (DZ, QKX),]; National Natural Science Foundation of China [Grants No. 52388201 (QKX), T2425009 (DZ), 12141402 (YZ), 12274249 (DZ), 12361141820 (DZ), 92477101 (YZ), 12504161 (HW)]; Innovation Program for Quantum Science and Technology [Grants No. 2021ZD0302600 (YZ), 2021ZD0302400 (DZ)].

## Author contributions

Y. Z. and H. W. fabricated the devices and carried out the transport measurements. Z. W. did the TEM studies. Y. Z., H. W., Z.-Y. S., Q.-K. X., and D.Z. analyzed the data and wrote the paper. All authors discussed the results and commented on the manuscript.

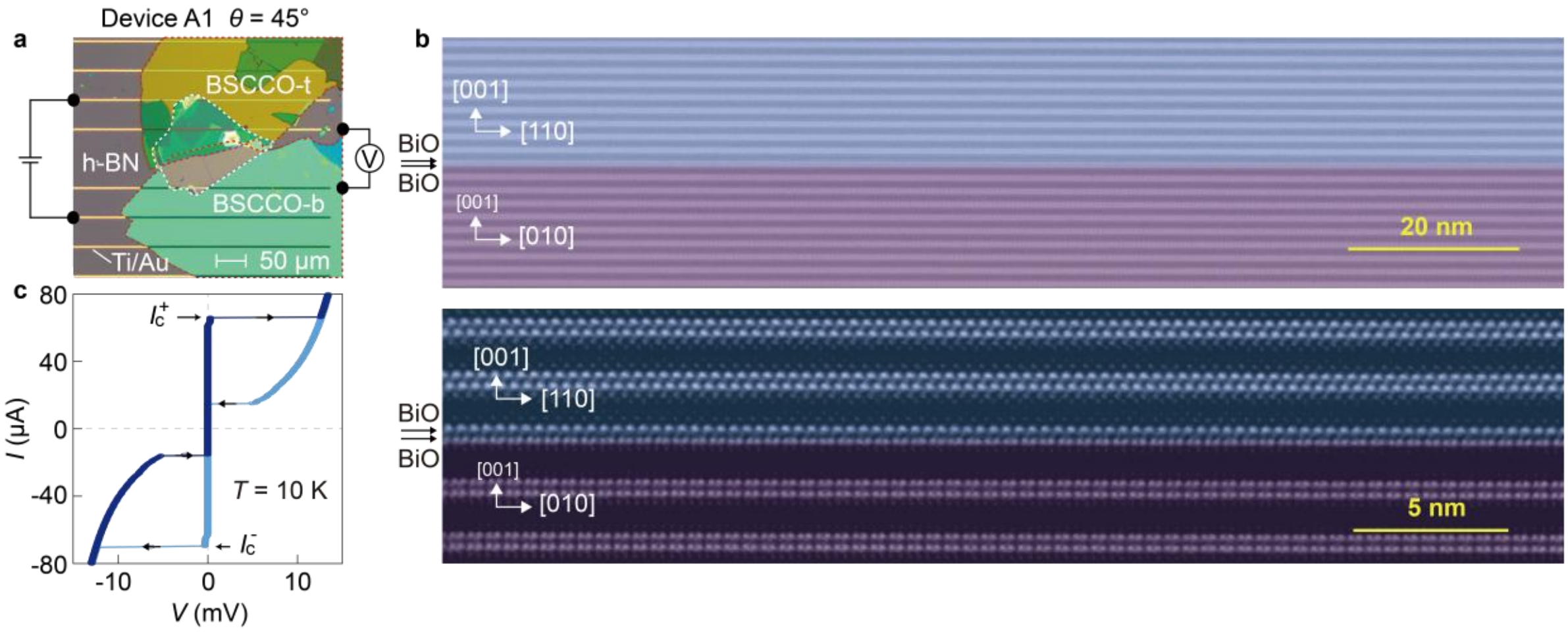

***Extended Data Figure 1: Device Image, I-V characteristic, and structural information. a,*** *Optical image of device A1.* ***b,*** *high-resolution cross-sectional TEM images of device A1. Blue and purple colors highlight the top and bottom flakes, respectively.* ***c,*** *I-V characteristic of device A1 without microwave irradiation. Arrows on the horizontal lines indicate the sweeping directions.* $I_c^{\pm}$ *marks the critical current.*

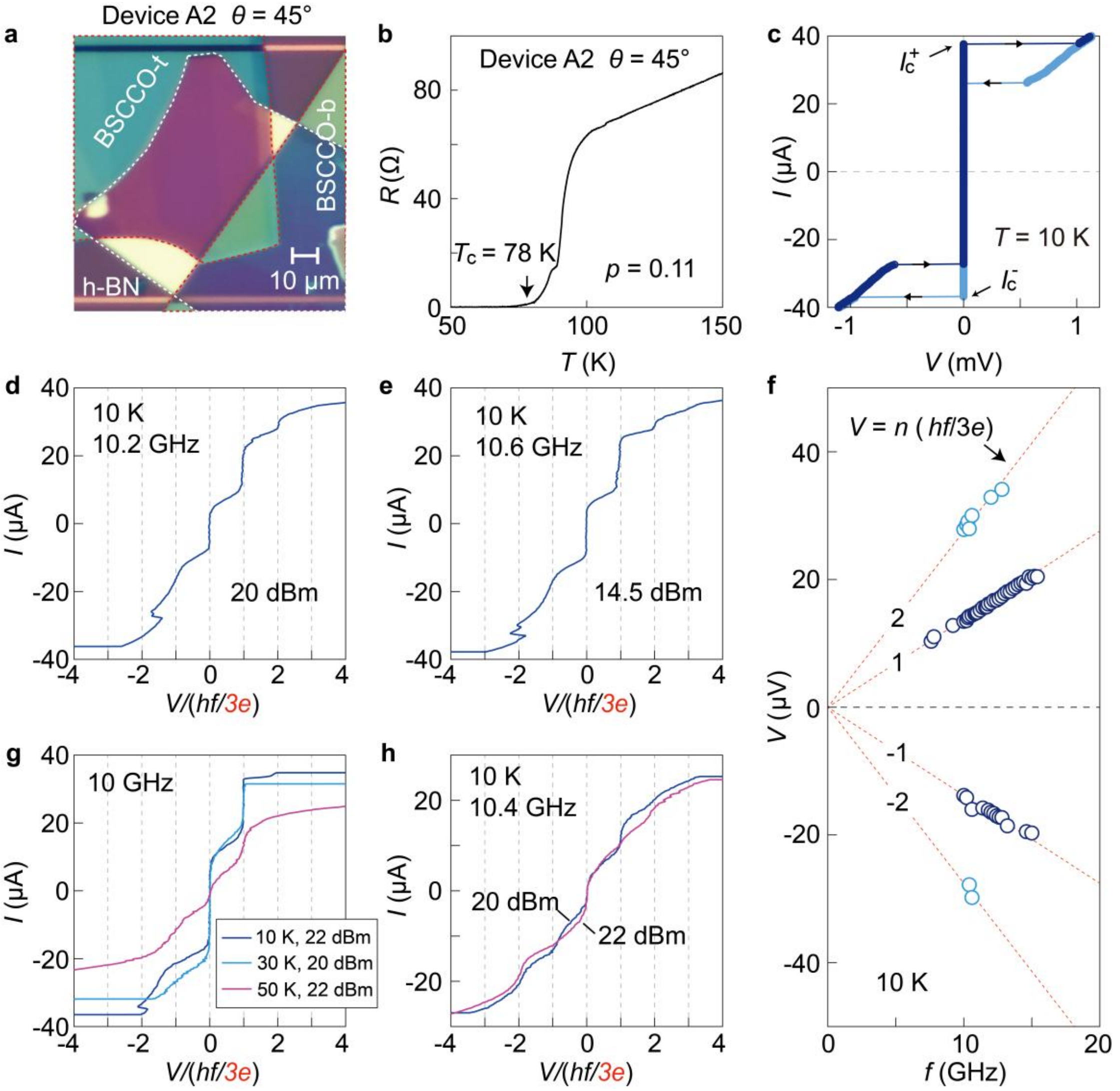

***Extended Data Figure 2: Charge-3e Josephson tunneling in device A2. a,*** *Optical image of the device A2.* ***b,*** *Temperature dependence of the junction resistance. The superconducting transition temperature* $T_c$ *and doping level* $p$ *are marked in the panel.* ***c,*** $I-V$ *characteristic measured at* $T$ *= 10 K.* $I_c^{\pm}$ *marks the positive and negative critical current.* ***d, e,*** $I$-$V$ *characteristics under microwave irradiation at frequencies of 10.2 GHz and 10.6 GHz, respectively. Here, the voltage values are normalized in units of* $hf/3e$. ***f,*** *Voltage positions of the Shapiro steps as a function of frequency* $f$ *for* $n = \pm 1$ *and* $\pm 2$. *The red dashed lines denote the theoretical relation* $V_s = nhf/3e$. ***g,*** $I$-$V$ *characteristics under microwave irradiation with a frequency of 10 GHz at temperatures from 10 K to 50 K.* ***h,*** $I$-$V$ *characteristics under different microwave powers and at a fixed frequency of 10.4 GHz.*

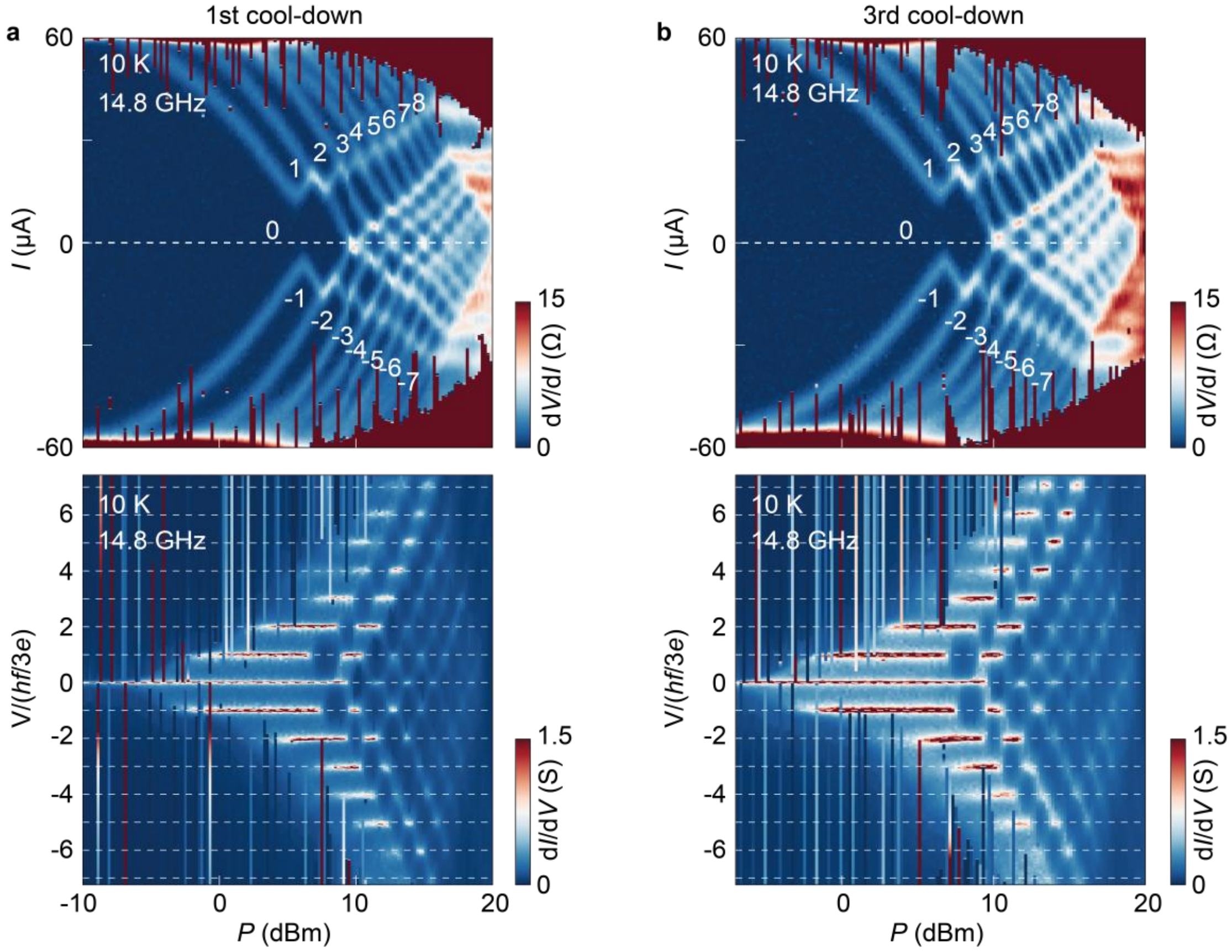

***Extended Data Figure 3: Unconventional Shapiros steps in two different thermal cycles. a,b*** *$dV/dI$ as a function of dc current $I$ and microwave power $P$ (top panels) and corresponding $dI/dV$ versus $V$ and $P$ (bottom panels) after the first and third cool-downs. The sample was warmed up to 120 K between the two cool-downs.*

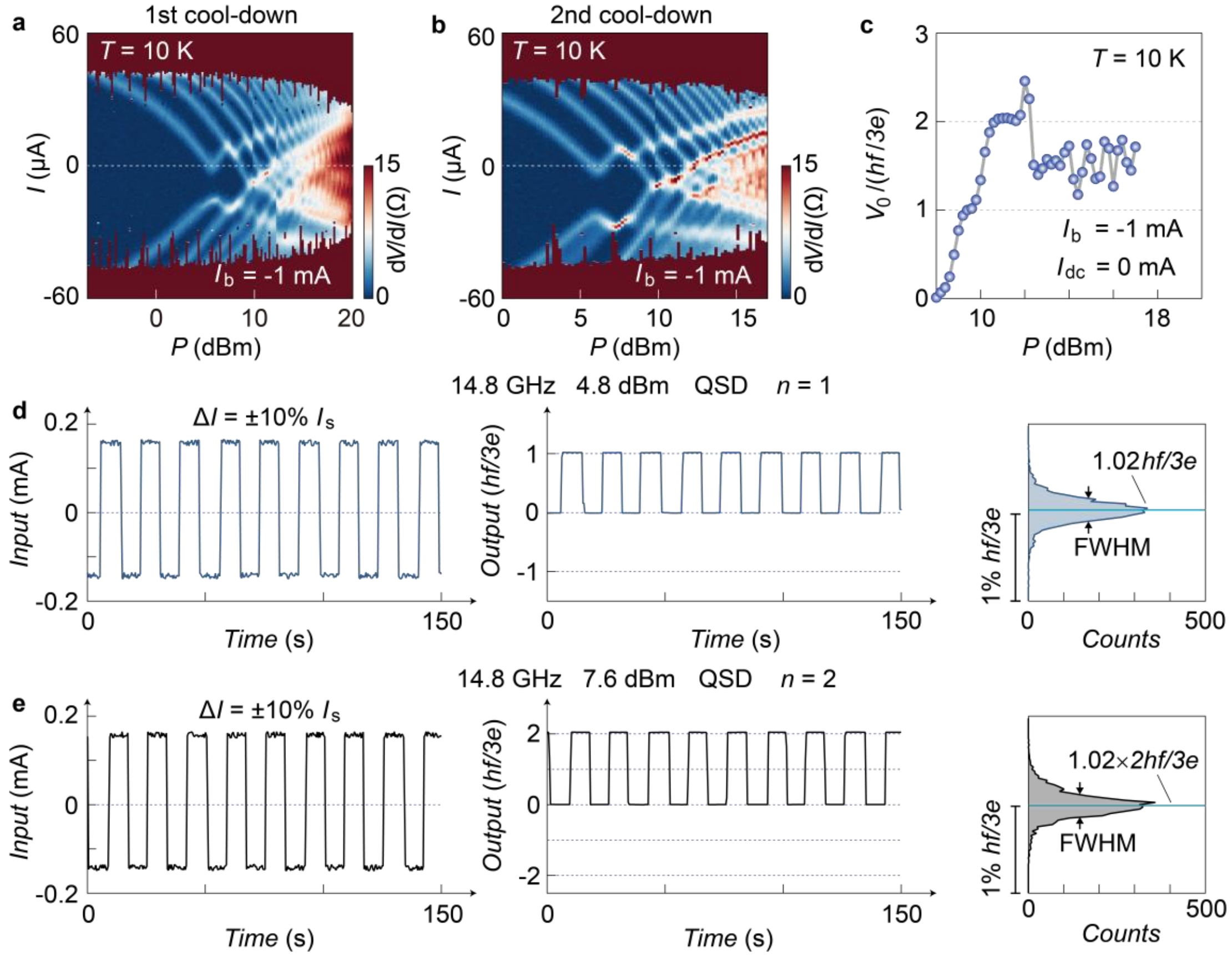

***Extended Data Figure 4: SD and QSD effects of device A1. a, b,*** *$dV/dI$ as a function of dc current $I$ and microwave power $P$ at 10 K after current annealing with -1 mA. Between the two cool-downs, the sample was warmed up to 120 K.* ***c,*** *Power dependence of the rectified voltage at zero current bias after inducing the SD effect with the current annealing protocol (-1 mA).* ***d, e,*** *Representative input waveforms (left), correspondent outputs (middle), and histogram of the output at $n$ = 1 (panel* ***d****) and 2 (panel* ***e****) in the QSD regime. Noise ($\Delta I$) is introduced to the input: ±10%$I_s$, where $I_s$ stays in the QSD regime. Each histogram in the right panel is taken for data over 200 repetitions. Vertical arrows indicate the full width at half maximum (FWHM). Solid horizontal line indicates the peak value of the output signal in half of the period.*

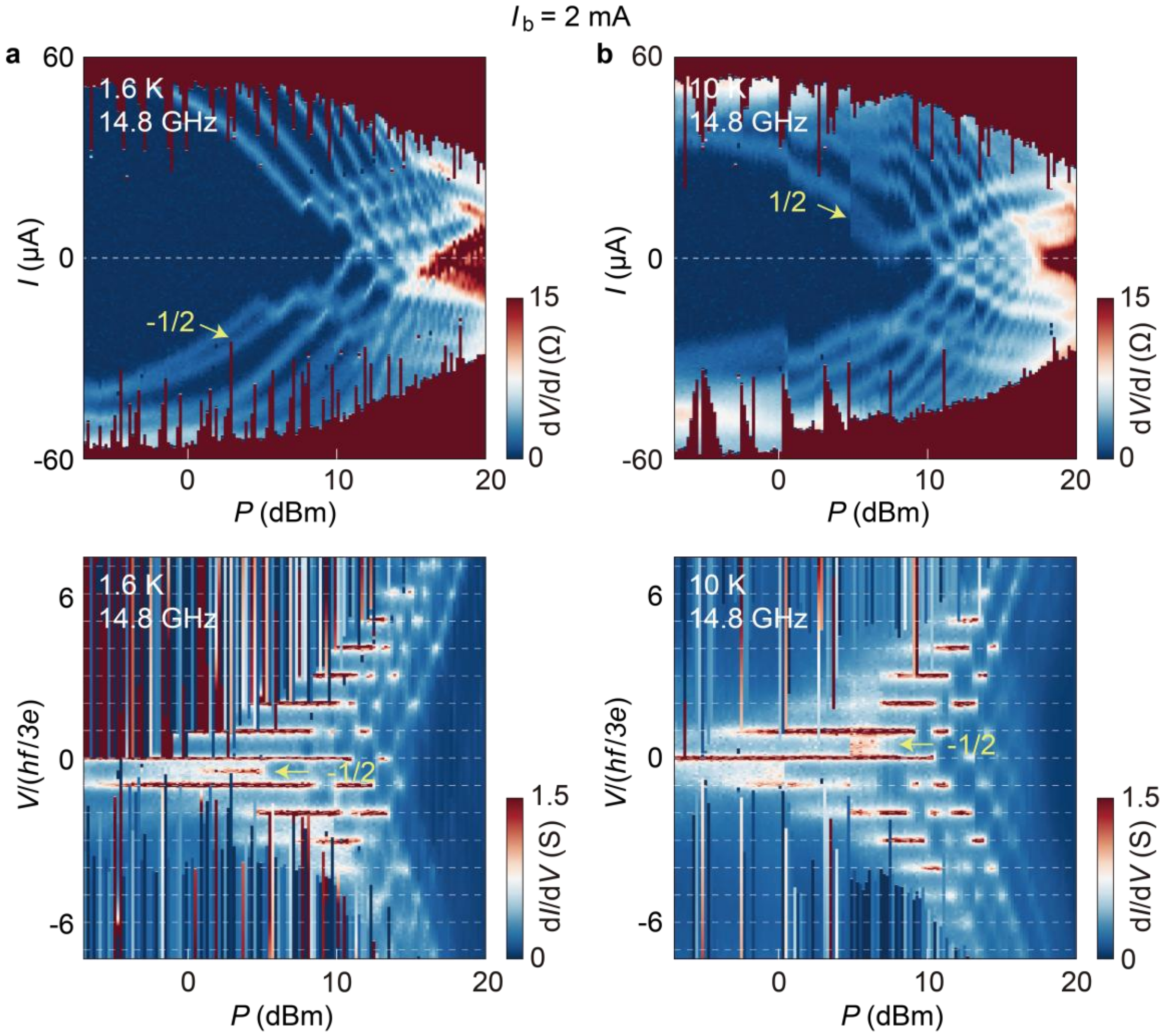

***Extended Data Figure 5: Half-integer Shapiro steps after applying bias current of 2 mA. a, b,*** *Differential resistance* $dV/dI$ *as a function of dc current* $I$ *and microwave power* $P$ *(Top panel) and corresponding* $dI/dV$ *versus normalized voltage* $V$ *and* $P$ *(bottom panel) at 1.6 K (panel* ***a****) and 10 K (panel* ***b****), respectively. Yellow arrows mark the half steps.*

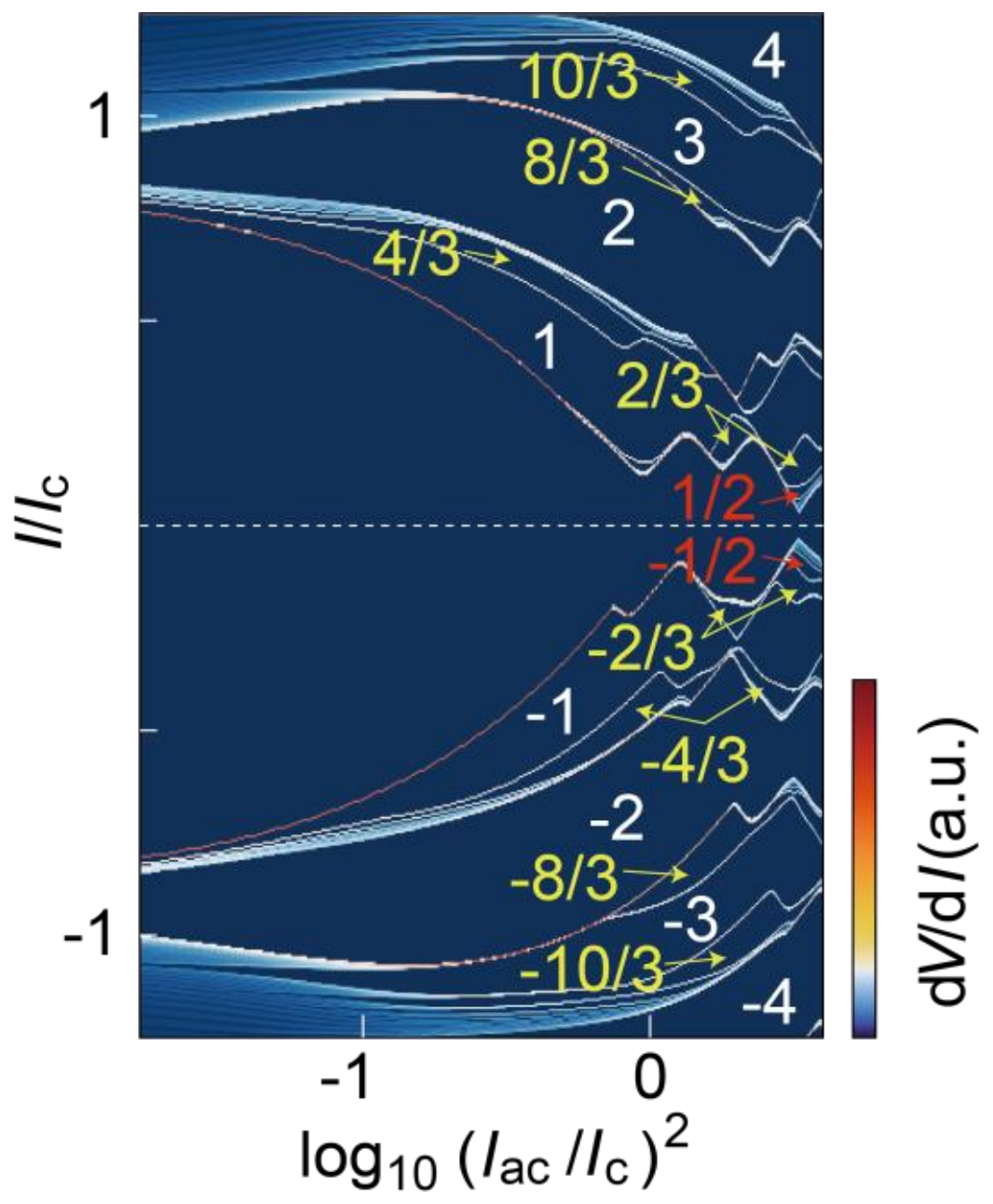

***Extended Data Figure 6:*** *Simulated evolution of Shapiro steps with* $I(\varphi) = I_{c1}sin(1.5\varphi) + I_{c2}sin(2\varphi) - I_m cos\varphi$. *For the particular simulation, we choose* $I_{c1} = I_{c2} = I_m = I_c/3$. *Numbers indicate the integral and fractional Shapiro steps in unit of* $hf/2e$.

| Device | Twist angle (°) | $T_c$ (K) | Area (μm$^2$) | $I_c$(μA) @10 K |
| --- | --- | --- | --- | --- |
| A1 | 45 | 77 | 1095 | 60 |
| A2 | 45 | 78 | 586 | 37 |
| B | 45 | 85 | 422 | 920 |

***Extended Data Table 1:*** *Comparison of the three junctions presented in the main text.*

| Device | Twist angle (°) | $T_c$ (K) | $hf/2e$ | $hf/3e$ |
| --- | --- | --- | --- | --- |
| A1 | 45 | 77 | | √ |
| A2 | 45 | 78 | | √ |
| B | 45 | 85 | √ | |
| B31205 | 45 | 86 | √ | |
| B50628 | 45 | 80 | √ | |
| B50919 | 45 | 87 | √ | |
| B50929 | 45 | 85 | √ | |
| B51014 | 45 | 87 | √ | |
| B51208 | 45 | 81 | √ | |
| B51215 | 45 | 87 | √ | |
| B51219 | 45 | 88 | √ | |
| B60331 | 40 | 87 | √ | |
| B60526 | 45 | 83 | √ | |
| B60915 | 45 | 82 | √ | |

***Extended Data Table 2:*** *Overview of the Shapiro sequence in twisted cuprate junctions.*